\documentclass[10pt]{article}

\usepackage{amsmath}
\usepackage{amssymb}

\usepackage{graphicx}

\usepackage{cite}

\usepackage{color}

\makeatletter
\renewcommand{\@biblabel}[1]{\quad#1.}
\makeatother

\date{}

\begin{document}

\begin{flushleft}
{\Large
\textbf{Foot Bone in Vivo: Its Center of Mass and Centroid of Shape}
}
\\
Yifang Fan$^{1,\ast}$,
Mushtaq Loan$^{2}$,
Yubo Fan$^{3}$,
Zhiyu Li$^{4}$,
Changsheng Lv$^{1}$
\\
\bf{1} Center for Scientific Research, Guangzhou Institute of Physical Education, Guangzhou 510500, P.R. China
\\
\bf{2} International School, Jinan University, Guangzhou 510632, P.R. China
\\
\bf{3} Bioengineering Department, Beijing University of Aeronautics and Astronautics, Beijing 100191, P.R. China
\\
\bf{4} College of Foreign Languages, Jinan University, Guangzhou 510632, P.R. China
\\
$\ast$ E-mail: tfyf@gipe.edu.cn
\end{flushleft}

\section*{Abstract}
This paper studies foot bone geometrical shape and its mass distribution and establishes an assessment method of bone strength. Using spiral CT scanning, with an accuracy of sub-millimeter, we analyze the data of 384 pieces of foot bones in vivo and investigate the relationship between the bone's external shape and internal structure. This analysis is explored on the bases of the bone's center of mass and its centroid of shape. We observe the phenomenon of superposition of center of mass and centroid of shape fairly precisely, indicating a possible appearance of biomechanical organism. We investigate two aspects of the geometrical shape, (i) distance between compact bone's centroid of shape and that of the bone and (ii) the mean radius of the same density bone issue relative to the bone's centroid of shape. These quantities are used to interpret the influence of different physical exercises imposed on bone strength, thereby contributing to an alternate assessment technique to bone strength.

\section*{Introduction}
The structure and function of bone largely depends on its mechanical and biological environments~\cite{Fung,Buckwalter,Rubin2001,Marco,Norbert}. When bone is bearing external force, bone tissue undergoes adaptive changes such as physiological activities of remodeling and reconstruction~\cite{Burger}. These changes emerge as those of external shape and internal structure so as to maximize its potential to bear external load~\cite{Harrigan1984,Bagge,Ruff}. Bone's adaptive changes to the external force are regarded as an optimization process~\cite{Huiskes,Harrigan,Marco} which observes some special law. Current research results, however, did not provide us with convincing explanations to this special law.

Bone strength is an important index to diagnose osteoporosis~\cite{Frost}. That justifies why bone strength has always been an issue in biomechanical research. Bone strength is determined by factors such as bone external shape and internal structure. How to measure bone shape and structure accurately largely depends on equipment. As a result, the advanced measurement equipment affects bone biomechanical research. With the development of CT scanning technology, more reliable modeling methods emerged. Many mechanical models, both macroscopically and microcosmically, depict the relationship between bone density, geometric shape, internal structure and bone strength~\cite{Gemunu,Crawford,Weinkamer,Rubin2005}. This has made a quantitative assessment of bone strength more reliable. But in clinical medicine, no other quantitative assessment index has been more widely used than bone mineral density, which actually cannot objectively reflect the real condition of bone strength~\cite{Crawford,Rubin2005,Bagi}. It is imperative to explore a quantitative and practical bone strength assessment index.

External force can change bone shape and structure~\cite{Doube} and in turn, bone shape and structure affect the result of external force. When we do research on the interaction of one body with another, bone's centroid of shape (COS) and center of mass (COM) have been considered as two basic physical quantities. Concerning the above-mentioned two issues, this paper focuses on selecting COS (shape) and COM (structure) to study the variation laws of bone in its adaptation process to external force and based upon these two quantities, we propose a couple of bone strength assessment indexes.

\section*{Materials and methods}
Previous studies~\cite{Cheng,Hudelmaier,Mittra} have shown that the isotropic materials' positions of COM and COS coincide since these materials form a closed three dimensional region. On the other hand, there exists enough experimental evidence that signatures the bone to be anisotropic ~\cite{Roesler,Turner1999}, in which case the COM and COS will superpose only under specific conditions. The broad understanding is still limited and suffers the lack of comprehensive theoretical model that shows a qualitative relationship between the position of COM and COS in anisotropic materials. The anisotropy of bone acoustic properties may also have a negative impact on imprecision. For example, the orientation of the calcaneus, not only around the leg axis but also around the heel-toe axis, must be controlled. Small movements of the leg or of the knee may slightly rotate the calcaneus around its long axis, resulting in undesired variations of the observables and positioning errors. Whereas several devices offer means of minimizing these variations, we choose CT scanning of foot bone to assess bone strength in vivo and examine the relationship between the positions of COM and COS in anisotropic materials.

The test equipment is Brilliance 64-slice Scanner by Philips, Netherlands, provided by Image Processing Center of Zhujiang Hospital. Scan settings are: frame bone tissue; power: 120kv; pixel size: 0.50mm; layer distance: 0.50mm. The scanning is conducted along both feet transect, from top to bottom.

Altogether, we have collected data of 384 pieces of bone - both from the volleyballers (with average height, weight and age of $183.94\pm3.90cm$, $69.80\pm5.20kg$ and $21.88\pm0.99 yrs$, respectively) and wrestlers (with average height, weight and age of $168.00\pm5.68cm$, $65.52\pm5.16kg$ and $21.00\pm2.78 yrs$, respectively), i.e. 32 pieces of 12 types of bones: calcaneus(Ca.), talus(Ta.), navicular(Na.), cuboid(Cu.), lateral cuneiform(La.), intermediate cuneiform(In.), medial cuneiform(Me.), first metatarsal(Fi.), second metatarsal(Se.), third metatarsal(Th.), fourth metatarsal(Fo.) and the fifth metatarsal(Fif.). The number of all of the differential elements of volume is 71,836,054. The subjects are male volleyball players from our institute and male wrestlers from Provincial Sports School. It has been confirmed before the test that every subject has been trained as a professional player for more than five years. Before the test, each subject's medical history was inquired and all the subjects were x-rayed to exclude subjects with diseases such as foot pathological change, deformity or injury to make sure that their physical conditions meet the requirements of the test. This study has been carried out according to the existing rules and regulations of our institute's Ethnic Committee.

\section*{Results}
\begin{figure}[!ht]
\begin{center}
\begin{tabular}{cccc}
 \includegraphics[width=12.8cm]{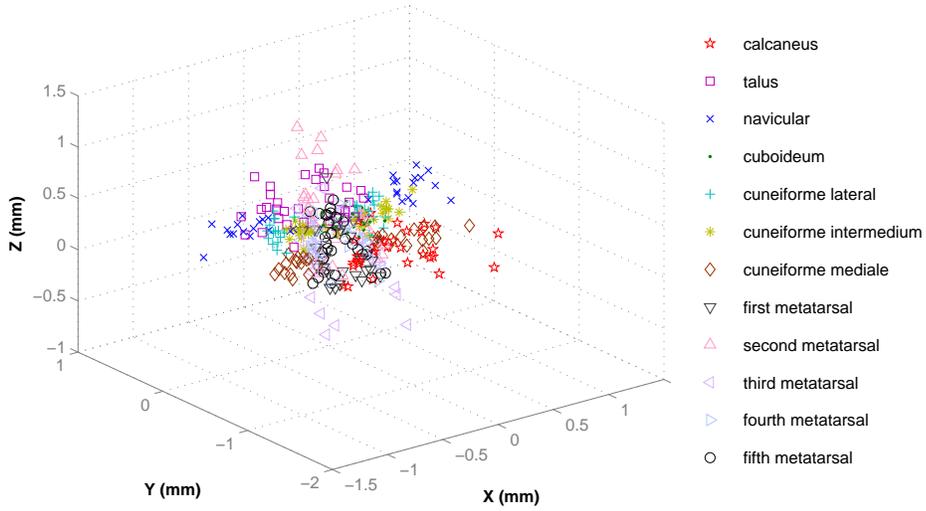}
\end{tabular}
\caption{\label{fig1} Relationship between the positions of foot bone's COM and COS.}
\end{center}
\end{figure}

Following~\cite{Ciarelli,Rho}, we separate foot bone to calculate the volume, surface area and average bone density (Table~\ref{t1}).
\begin{table}
\caption{\label{t1}Foot bone volume, surface area and bone density (Mean$\pm$SD)}
\small
\begin{center}
\item[]\begin{tabular}{@{}*{7}{l}}
\hline
&\textbf{Wrestler}& & &\textbf{Volleyballer}& &\\
&Volume$(cm^{3})$&Area$(cm^{2})$&Density$(g/cm^{3})$&Volume$(cm^{3})$&Area$(cm^{2})$&Density$(g/cm^{3})$\\
\hline
Ca.&$71.01\pm8.46$&$107.39\pm8.83$&$1.47\pm0.04$&$83.94\pm6.05$&$120.70\pm5.56$&$1.49\pm0.05$\\
Ta.&$38.30\pm4.33$&$71.38\pm5.41$&$1.63\pm0.04$&$43.87\pm3.33$&$80.11\pm5.97$&$1.65\pm0.04$\\
Na.&$11.45\pm1.39$&$31.21\pm2.73$&$1.56\pm0.04$&$13.44\pm1.51$&$34.78\pm3.00$&$1.58\pm0.05$\\
Cu.&$13.87\pm1.61$&$33.14\pm2.77$&$1.46\pm0.04$&$15.09\pm2.69$&$35.24\pm4.78$&$1.47\pm0.05$\\
La.&$5.91\pm0.69$&$19.09\pm1.56$&$1.51\pm0.04$&$6.79\pm0.61$&$20.99\pm1.28$&$1.53\pm0.06$\\
In.&$4.43\pm0.66$&$15.69\pm1.56$&$1.59\pm0.04$&$5.20\pm0.44$&$17.56\pm1.00$&$1.64\pm0.06$\\
Me.&$10.76\pm1.48$&$28.60\pm2.73$&$1.52\pm0.03$&$12.20\pm1.04$&$31.02\pm1.91$&$1.58\pm0.05$\\
Fi.&$16.94\pm2.23$&$44.90\pm3.89$&$1.62\pm0.05$&$20.93\pm2.25$&$51.94\pm3.47$&$1.65\pm0.05$\\
Se.&$9.01\pm1.29$&$33.72\pm3.25$&$1.73\pm0.07$&$11.65\pm0.77$&$40.56\pm2.08$&$1.76\pm0.08$\\
Th.&$7.72\pm0.58$&$30.23\pm1.60$&$1.70\pm0.05$&$8.99\pm1.07$&$34.50\pm2.61$&$1.68\pm0.07$\\
Fo.&$7.47\pm0.78$&$28.80\pm2.19$&$1.66\pm0.04$&$8.88\pm0.92$&$32.97\pm2.05$&$1.66\pm0.05$\\
Fif.&$8.83\pm1.09$&$30.92\pm2.64$&$1.72\pm0.05$&$9.65\pm1.07$&$33.73\pm2.49$&$1.71\pm0.05$\\
\hline
\end{tabular}
\end{center}
\end{table}
Due to their significant differences in height and weight, there exist statistically significant differences between the volleyballers' and wrestlers' foot bone volume and surface area. However, the acquired differential element of volume ($dv$) is a constant.
To calculate bone's COM and COS, we consider n number of infinitesimal volume elements dv with density $\rho$. Choosing the orthogonal coordinate system to locate the position of elementary volume dv, we use
\begin{displaymath}
x_{p}=\frac{\sum x\rho}{\sum\rho}, \\
y_{p}=\frac{\sum y\rho}{\sum\rho}, \\
z_{p}=\frac{\sum z\rho}{\sum\rho}
\end{displaymath}
and
\begin{displaymath}
x_{c}=\frac{\sum x}{\sum\rho}, \\
y_{c}=\frac{\sum y}{\sum\rho}, \\
z_{c}=\frac{\sum z}{\sum\rho}
\end{displaymath}
to evaluate the COM and COS of the bone, respectively. Setting the COS as the origin, the position of the bone's COM relative to the COS can be calculated by calculating the spatial separation between any two points in three dimensions (Fig.~\ref{fig1}). We notice that about $99.22\%$; $99.22\%$ and $99.74\%$ of the distance between the foot bone COM and COS is within the range of $\pm1mm$ from X; Y and Z axes, respectively. With the pixel size of spiral CT scanning at 0.50mm and its layer distance at 0.50mm, the measured spatial parameters suggest a superposition of the foot bone's COM and COS, which is collected and displayed in Fig.~\ref{fig1}.

\begin{figure}[!ht]
\begin{center}
\begin{tabular}{cccc}
 \includegraphics[width=12.8cm]{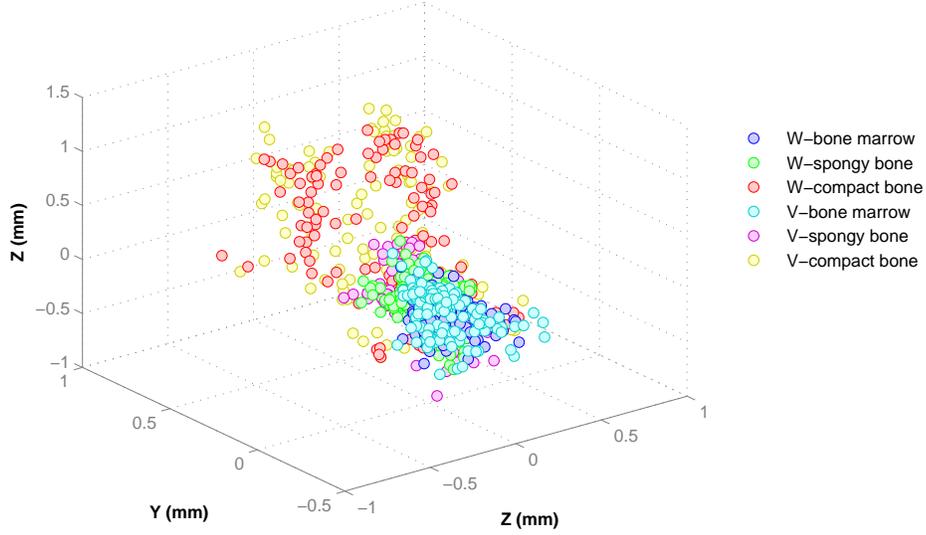}
\end{tabular}
\caption{\label{fig2} Relationship between COM and COS of tissues of marrow, spongy bone and compact bone (W refers to wrestlers and V volleyballers.)}
\end{center}
\end{figure}

To explore this phenomenon of superposition in tissues, we separate the bone in terms of marrow, spongy bone and compact bone by applying the CT value of the corresponding bone tissue relationship ~\cite{Rho,Taylor2002}. Following the above procedure, we calculate the COM and COS of three tissues of each bone. The corresponding relationship is shown in Fig.~\ref{fig2}.
A similar behaviour of the distance between tissues' COM and COS is observed indicating the presence of superposition of the positions of COM and COS of tissues of marrow, spongy bone and compact bone. Whether the superposition observed from the above analysis is a genuine destined phenomenon or an accidental event can be resolved from what follows.

\section*{Discussion}
Having explored the superposition, the results can be used to extract the biomechanical significance
of this principle. To this end we conduct an analysis of the moment of inertia, a physical quantity which
is related both to the bone's geometric shape and its mass distribution. Using the optimal function

\begin{equation}\label{eq3}
\min\psi(p_{c})=\sum m_{i}\left(\left( x_{i}-x_{b}\right)^{2}+\left( y_{i}-y_{b}\right)^{2}+\left( z_{i}-z_{b}\right)^{2}\right),
\end{equation}
where $p_{b}(x_{b},y_{b},z_{b})$ and $p_{i}(x_{i},y_{i},z_{i})$ refer to the positions of bone's COM and any position of dv relative
to CT image coordinate, we differentiate the right side of Eq.~\ref{eq3} to obtain
\begin{displaymath}
\sum 2 m_{i}\left(\left|x_{i}-x_{b}\right|+\left|y_{i}-y_{b}\right|+\left|z_{i}-z_{b}\right|\right).
\end{displaymath}
For a closed and continuous three-dimensional geometric body, the series
\begin{displaymath}
\sum m_{i},\sum \left(\left|x_{i}-x_{b}\right|+\left|y_{i}-y_{b}\right|+\left|z_{i}-z_{b}\right|\right),\sum m_{i}\left(\left|x_{i}-x_{b}\right|+\left|y_{i}-y_{b}\right|+\left|z_{i}-z_{b}\right|\right),
\end{displaymath}
converge. Thus in the limit $m_{i}\rightarrow0$, using the Abelian Theorem [31], the above equation becomes
\begin{displaymath}
\sum m_{i}\left(\left| x_{i}-x_{b}\right|+\left| y_{i}-y_{b}\right|+\left|z_{i}-z_{b}\right|\right)=\sum m_{i}\sum \left(\left|x_{i}-x_{b}\right|+\left|y_{i}-y_{b}\right|+\left|z_{i}-z_{b}\right|\right).
\end{displaymath}
Suppose
\begin{displaymath}
\sum m_{i}\sum \left(\left|x_{i}-x_{b}\right|+\left|y_{i}-y_{b}\right|+\left|z_{i}-z_{b}\right|\right)=0,
\end{displaymath}
then according to Parallel Axis Theorem, the point $(x_{b},y_{b},z_{b})$ must be the COM, and
\begin{displaymath}
\sum \left|x_{i}-x_{b}\right|=0, \\
\sum \left|y_{i}-y_{b}\right|=0, \\
\sum \left|z_{i}-z_{b}\right|=0.
\end{displaymath}
Since
\begin{displaymath}
\sum x_{b}=nx_{b}, \\
\sum y_{b}=ny_{b}, \\
\sum z_{b}=nz_{b},
\end{displaymath}
therefore
\begin{displaymath}
x_{b}=x_{c}, y_{b}=y_{c}, z_{b}=z_{c}.
\end{displaymath}
This implies that
\begin{equation}\label{eq4}
p_{c}=p_{b}
\end{equation}

Eq.~\ref{eq4} suggests that the optimal program for the heterogeneous bone tissue mass distribution should be the superposition of its COM and COS. This coincidence can avoid the emergence of eccentric force, enhance the bone's potential to bear the external force, optimize the bone's internal structure and then achieve the goal to use the minimal material to maximize its function~\cite{Roesler}. Consequently, in order to get adapted to the mechanical environment, the bone undergoes continuous functional adaptive changes~\cite{Marco,Huiskes, Harrigan}. In this process, the bone always observes the optimized principle of superposition between the bone's COM and its COS.

\begin{figure}[!ht]
\begin{center}
\begin{tabular}{cccc}
 \includegraphics[width=12.8cm]{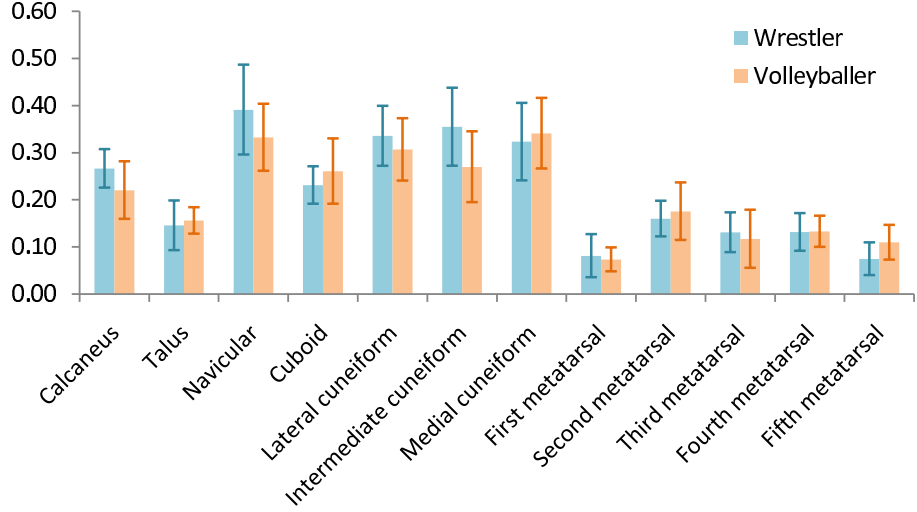}
\end{tabular}
\caption{\label{fig3} Relationship between the compact bone's COS relative to the bone's COS. Calculate each individual bone's compact bone COS and the bone's COS. Calculate the distance between the compact bone's COS and the bone's COS. Apply $\bar{r}=\frac{1}{n}\sum^{n}_{1}\left( \sqrt{(x_{i}-x_{c})^{2}+(y_{i}-y_{c})^{2}+(z_{i}-z_{c})^{2}}\right)$ to calculate all bone tissues' mean distribution radii relative to the bone's COS and use $d_{compact}/\bar{r}$ to standardize the distance between the compact bone's COS and the bone's COS.}
\end{center}
\end{figure}

Experimental results show that the strength of compact bone of bone in vitro is many times more than that of the spongy bone~\cite{Evans1957,Currey}. That means that the bone strength is mainly determined by compact bone. Fig.~\ref{fig2} indicates that the COM and COS of compact bone are in accordance with Eq.~\ref{eq4}. The position of compact bone's COS relative to the bone's COS will affect the external force. However, will the external force change the relationship of their positions? We calculate the position of compact bone's COS relative to the bone's COS. Set the coordinate origin at the bone's COS, the relationship of positions of the compact bone's COS relative to the bone's COS could be gained. See Fig.~\ref{fig3}.

\begin{figure}[!ht]
\begin{center}
\begin{tabular}{cccc}
 \includegraphics[width=12.8cm]{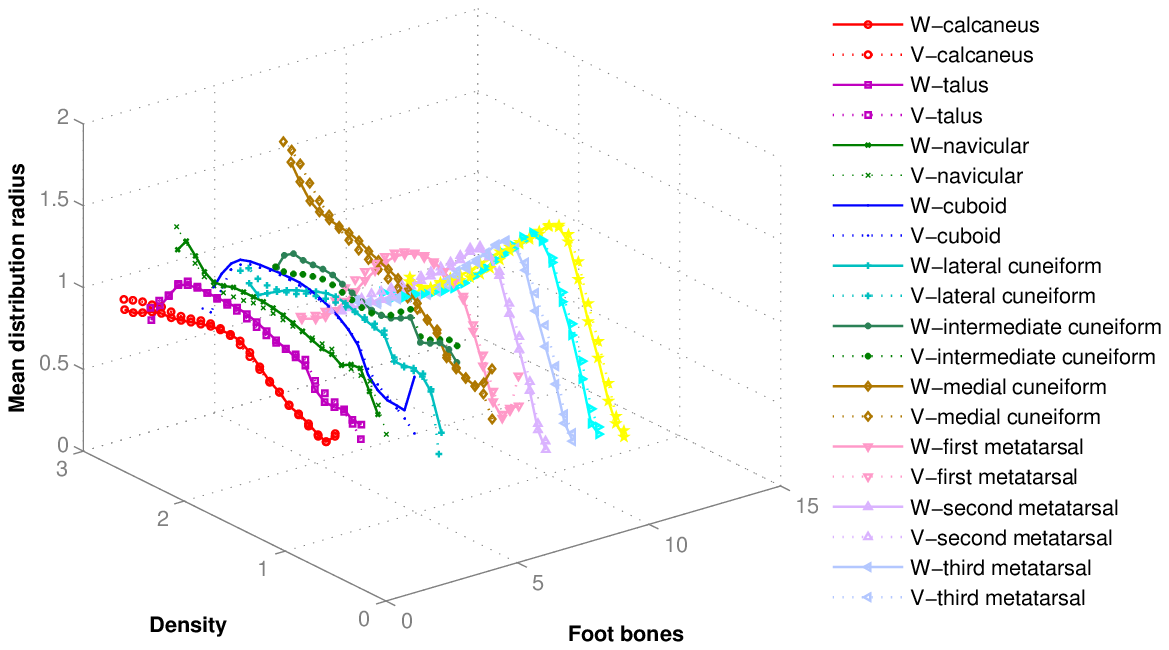}
\end{tabular}
\caption{\label{fig4} Relationship between the bone's density and the MDR relative to the bone's COS. We calculate the MDR to the bone's COS by applying $\bar{r_{j}}=\frac{1}{n_{j}}\sum^{n_{j}}_{1}\left( \sqrt{(x_{i}-x_{c})^{2}+(y_{i}-y_{c})^{2}+(z_{i}-z_{c})^{2}}\right)$, calculate the MDR of the bone tissue relative to bone's COS by $\bar{r}=\frac{1}{n}\sum^{n}_{1}\left( \sqrt{(x_{i}-x_{c})^{2}+(y_{i}-y_{c})^{2}+(z_{i}-z_{c})^{2}}\right)$ and then standardize the MDR relative to the bone's COS by $\bar{r_{j}/\bar{r}}$.}
\end{center}
\end{figure}

When the compact bone's COS moves towards the bone's COS, what happens to the mass distribution of bone tissue? In order to explore this issue, we propose a concept of mean distribution radius (MDR) of same density tissue relative to its bone's COS and explore the relationship between the MDR and the density. The relationship between twelve types of foot bone density and the MDR relative to the bone's COS is shown in Fig.~\ref{fig4}.

When a volleyballer takes off to spike, the braking movement has a great impact on the calcaneus. In Table~\ref{t1}, the volleyballers' calcaneus density is larger than that of the wrestlers. But there is no statistically significant difference ($p=0.1887$). In Fig.~\ref{fig3}, the distance of the volleyballers' calcaneus compact bone's COS to the bone's COS is shorter than that of the wrestlers and it has a statistically significant difference ($p=0.0178$). In Fig.~\ref{fig4}, the MDR of the volleyballers' calcaneous relative to the bone's COS begins to be larger than that of the wrestlers from the density of compact bone on; especially when comparing this with the results from the marrow and spongy bone tissues, this difference is outstanding. The wrestlers' fierce body combats carry great strength to their fifth metatarsal from the front, rear, left and right. Table~\ref{t1} suggests the wrestlers fifth metatarsal density is greater than that of the volleyballers, but again, there is no statistically significant difference ($p=0.4231$). In Fig.~\ref{fig3}, the distance of the wrestlers' fifth metatarsal compact bone COS to the bone's COS is shorter than that of the volleyballers and there is a big difference ($p=0.0098$). In Fig.~\ref{fig4}, the MDR of the wrestlers' fifth metatarsal relative to the bone's COS begins to become bigger from the density of compact bone on than that of the volleyballers. The difference grows especially when approaching their peak values. In both sport events, the impact strength to the first metatarsal is obvious. From Table~\ref{t1}, we can see that in the similar morphological first to fifth metatarsal, the lowest density goes to the first metatarsal, which does not sound very reasonable. (Factors such as volume and joint segmental area might have affected bone density, which might as well mean the limitation of bone density assessment index.) In Fig.~\ref{fig3}, the distance of both athletic groups' first metatarsal compact bone COS to the bone's COS is the shortest. In Fig.~\ref{fig4}, comparing with the second to the fourth metatarsal, the MDR of both athletic groups' first metatarsal relative to the bone's COS begins to enlarge.

The above results and analyses suggest that greater strength sports activities can enlarge the bone density peak values~\cite{Grimston,Heaney}. The idea that the impact strength can add bone mass significantly~\cite{Calbet,Turner2003,Torstveit} has been verified by the volleyballers' and wrestlers' foot bones. This is the result of bone's adaptation towards mechanical environment. When assessing the above results by the two indices, i.e. the distance between the compact bone's COS and the bone's COS and the MDR relative to the bone's COS, the results are consistent with the foot bone feature of load from both athletic groups.

\section*{Conclusion}
This research studies the relationship between the positions of the COM and COS of foot bone in vivo by an analysis of sub-millimeter accuracy. The phenomenon of superposition of foot bone's COM and COS has been discovered. This phenomenon does not exist only in the relationship between the positions of bone's COM and COS but also in the relationship between the COM and COS of tissues such as marrow, spongy bone and compact bone. We establish the relationship between the tissues' masses and their positions, with which we analyze and illustrate that the phenomenon of superposition of bone's COM and COS is caused by the approximate symmetry of distribution of the bone tissue relative to the bone's COS. We define an active function and determine an optimal condition and prove the superposition of foot bone's COM and COS to be an optimized distribution program.

We propose the concept of distance between the compact bone's COS and its bone's COS and discover the relationship between the distance (of the compact bone's COS and its bone's COS) and the loading type. This relationship is represented by the phenomenon that the impact strength has made the compact bone's COS move towards the bone's COS. This movement symbolizes a functional adaptation of bone in its structure. The physiological activity of the middle aged and seniors is mostly reconstruction~\cite{Taylor}. When their bone masses are gradually decreasing, it is essential to look into the possibility of whether physical exercises can diminish the bone loss and change the movement's direction. This is meaningful and worthwhile research.

We have studied the relationship between the bone's density and the MDR of foot bone relative to the bone's COS and we have observed the obvious difference that various exercises exert on compact bone's MDR. Since MDR has much to do with bone strength, the MDR relative to the bone's COS and the distance between compact bone's COS and the bone's COS can assess the bone strength more reliably.

In conclusion, in the physiological activities of foot bones' adaptation to their external mechanical environment, they always observe the optimized distribution law of superposition between the COM and COS. By changing the distance between the compact bone's COS and the bone's COS and by changing bone tissue MDR to change bone strength, the bone can adapt to meet the requirements of muscle strength and training load to ensure the normal activities. The movement of compact bone's COS and the change of MDR all signal the adaptation of bone.

\section*{Acknowledgments}

This project was funded by National Natural Science Foundation of China under the grant number of $10772053$, $10972061$ and by Key Project of Natural Science Research of Guangdong Higher Education Grant No $06Z019$. The authors would like to acknowledge the support from Image Center of Zhujiang Hospital and Clinic of Guangzhou Institute of Physical Education.


\begin{thebibliography}{10}
\bibitem{Fung} Fung Y C (1990) \emph{Biomechanics: Motion, Flow, Stress, and Growth.} New York:
Springer, pp499-532
\bibitem{Buckwalter} Buckwalter JA, Glimcher MJ, Cooper RR, Recker R (1996) Bone biology. I: Structure, blood supply, cells, matrix, and mineralization. Instr Course Lect 45: 371-386.
\bibitem{Rubin2001} Rubin C, Turner AS, Bain S, Mallinckrodt C, McLeod K (2001) Anabolism - Low mechanical signals strengthen long bones. Nature 412: 603-604.
\bibitem{Marco} Rusconi M, Zaikin A, Marwan N, Kurths J (2008) Effect of stochastic resonance on bone loss in osteopenic conditions. Phys Rev Lett 100 128101.
\bibitem{Norbert} Marwan N, Kurths J, Thomsen JS, Felsenberg D, Saparin P (2009) Three-dimensional quantification of structures in trabecular bone using measures of complexity. Phys Rev E 79 021903.
\bibitem{Burger} Burger EH, Klein-Nulend J (1999) Mechanotransduction in bone - role of the lacuno-canalicular network. Faseb Journal 13: S101-S112.
\bibitem{Harrigan1984} Harrigan TP and Mann RW (1984) Characterization of microstructural anisotropy in orthotropic materials using a second rank tensor. J Mater Sci 19: 761-767.
\bibitem{Bagge} Bagge M (2000) A model of bone adaptation as an optimization process. J Biomech 33: 1349-1357.
\bibitem{Ruff} Ruff C (2003) Ontogenetic adaptation to bipedalism: age changes in femoral to humeral length and strength proportions in humans, with a comparison to baboons. J Hum Evol 45: 317-349.
\bibitem{Huiskes} Huiskes R, Hollister SJ (1993) From Structure To Process, From Organ to Cell - Recent Developments of Fe-Analysis in Orthopedic Biomechanics. J Biomech Eng-T ASME 115: 520-527.
\bibitem{Harrigan} Harrigan TP, Hamilton JJ (1994) Bone Remodeling and Structural Optimization. J Biomech 27: 323-328.
\bibitem{Frost} Frost HM (1997) Defining osteopenias and osteoporoses: Another view (with insights from a new paradigm). Bone 20: 385-391.
\bibitem{Gemunu} Gunaratne GH, Rajapaksa CS, Bassler KE, Mohanty KK, Wimalawansa SJ (2002) Model for bone strength and osteoporotic fractures. Phys Rev Lett 88 068101.
\bibitem{Crawford}Crawford RP, Rosenberg WS, Keaveny TM (2003) Quantitative computed tomography-based finite element models of the human lumbar vertebral body: Effect of element size on stiffness, damage, and fracture strength predictions. J Biomech Eng-T ASME 125: 434-438.
\bibitem{Weinkamer} Weinkamer R, Hartmann MA, Brechet Y, Fratzl P (2004) Stochastic lattice model for bone remodeling and aging. Phys Rev Lett 93: 228102.
\bibitem{Rubin2005} Rubin CD (2005) Emerging concepts in osteoporosis and bone strength. Curr Med Res Opin 21: 1049-1056.
\bibitem{Bagi} Bagi CM, Hanson N, Andresen C, Pero R, Lariviere R, et al. (2006) The use of micro-CT to evaluate cortical bone geometry and strength in nude rats: Correlation with mechanical testing, pQCT and DXA. Bone 38: 136-144.
\bibitem{Doube} Doube M, Conroy AW, Christiansen P, Hutchinson JR, Shefelbine S (2009) Three-Dimensional Geometric Analysis of Felid Limb Bone Allometry. PLoS ONE 4(3): e4742.

\bibitem{Cheng} Cheng S, Sipila S, Taaffe DR, Puolakka J, Suominen H (2002) Change in bone mass distribution induced by hormone replacement therapy and high-impact physical exercise in post-menopausal women. Bone 31: 126-135.
\bibitem{Hudelmaier} Hudelmaier M, Kollstedt A, Lochmuller EM, Kuhn V, Eckstein F, et al. (2005) Gender differences in trabecular bone architecture of the distal radius assessed with magnetic resonance imaging and implications for mechanical competence. Osteoporosis Int 16: 1124-1133.
\bibitem{Mittra} Mittra E, Rubin C, Gruber B, Qin YX (2008) Evaluation of trabecular mechanical and microstructural properties in human calcaneal bone of advanced age using mechanical testing, mu CT, and DXA. J Biomech 41: 368-375.
\bibitem{Roesler} Roesler H (1987) The History of Some Fundamental-Concepts in Bone Biomechanics. J Biomech 20: 1025-1034.
\bibitem{Turner1999} Turner CH (1999) Toward a mathematical description of bone biology: The principle of cellular accommodation. Calcified Tissue Int 65: 466-471.
\bibitem{Ciarelli} Ciarelli MJ, Goldstein SA, Kuhn JL, Cody DD, Brown MB (1991) Evaluation of Orthogonal Mechanical-Properties and Density of Human Trabecular Bone From the Major Metaphyseal Regions With Materials Testing and Computed-Tomography. J Orthop R 9: 674-682.
\bibitem{Rho} Rho JY, Hobatho MC, Ashman RB (1995) Relations of Mechanical-Properties to Density and Ct Numbers in Human Bone. Med Eng Phys 17: 347-355.
\bibitem{Taylor2002} Taylor WR, Roland E, Ploeg H, Hertig D, Klabunde R, et al. (2002) Determination of orthotropic bone elastic constants using FEA and modal analysis. J Biomech 35: 767-773.
\bibitem{Hardy}Hardy G H (2002) \emph{A Course of Pure Mathematics.} 10th Ed. New York: Cambridge University Press
\bibitem{Evans1957} Evans FG, Lebow M (1957) Strength of Human Compact Bone Under Repetitive Loading. J Appl Physiol 10: 127-130.
\bibitem{Currey} Currey JD (1999) What determines the bending strength of compact bone? J Exp Biol 202: 2495-2503.
\bibitem{Grimston} Grimston SK and Zernicke RF (1993). Exercise related stress responses in bone. J Appl Biomech, 9: 2-14
\bibitem{Heaney} Heaney RP, Abrams S, Dawson-Hughes B, Looker A, Marcus R, et al. (2000) Peak bone mass. Osteoporosis Int 11: 985-1009.
\bibitem{Calbet} Calbet JAL, Dorado C, Diaz-Herrera P, Rodriguez-Rodriguez LP (2001) High femoral bone mineral content and density in male football (soccer) players. Med Sci Sport Exer 33: 1682-1687.
\bibitem{Turner2003}Turner CH, Robling AG (2003) Designing exercise regimens to increase bone strength. Exerc Sport Sci Rev 31: 45-50.
\bibitem{Torstveit}Torstveit MK, Sundgot-Borgen J (2005) Low bone mineral density is two to three times more prevalent in non-athletic premenopausal women than in elite athletes: a comprehensive controlled study. Brit J Sport Med 39: 282-287.
\bibitem{Taylor} Taylor D, Hazenberg JG, Lee TC (2007) Living with cracks: Damage and repair in human bone. Nat Mater 6: 263-268.
\end{thebibliography}
\end{document}